\documentclass[a4paper,fleqn]{cas-dc}
\usepackage[numbers]{natbib}
\usepackage[most]{tcolorbox}
\usepackage{dirtytalk}
\usepackage{fontawesome}
\usepackage{pifont}
\usepackage{url}
\usepackage{balance}
\usepackage[T1]{fontenc}
\usepackage{ragged2e}
\usepackage{makeidx}
\usepackage{orcidlink}

\definecolor{anti-flashwhite}{rgb}{0.95, 0.95, 0.96}
\definecolor{beige}{rgb}{0.96, 0.96, 0.86}
\definecolor{floralwhite}{rgb}{1.0, 0.98, 0.94}
\definecolor{gainsboro}{rgb}{0.86, 0.86, 0.86}
\definecolor{ghostwhite}{rgb}{0.97, 0.97, 1.0}
\definecolor{honeydew}{rgb}{0.94, 1.0, 0.94}
\definecolor{isabelline}{rgb}{0.96, 0.94, 0.93}
\definecolor{ivory}{rgb}{1.0, 1.0, 0.94}
\definecolor{magnolia}{rgb}{0.97, 0.96, 1.0}
\definecolor{mintcream}{rgb}{0.96, 1.0, 0.98}
\definecolor{pearl}{rgb}{0.94, 0.92, 0.84}
\definecolor{whitesmoke}{rgb}{0.90, 0.90, 0.90}

\def\tsc#1{\csdef{#1}{\textsc{\lowercase{#1}}\xspace}}
\tsc{WGM}
\tsc{QE}
\tsc{EP}
\tsc{PMS}
\tsc{BEC}
\tsc{DE}

\begin{document}
\let\printorcid\relax       
\let\WriteBookmarks\relax
\def\floatpagepagefraction{1}
\def\textpagefraction{.001}
\shorttitle{Warnings: Violation Symptoms Indicating Architecture Erosion}
\shortauthors{\textit{R. Li et al}.}
\title [mode = title]{Warnings: Violation Symptoms Indicating Architecture Erosion}

\author[1,2]{Ruiyin Li~\orcidlink{0000-0001-8536-4935}}
\ead{ryli_cs@whu.edu.cn}

\author[1]{Peng Liang~\orcidlink{0000-0002-2056-5346}}
\cormark[1]
\ead{liangp@whu.edu.cn}

\author[2]{Paris Avgeriou~\orcidlink{0000-0002-7101-0754}}
\ead{p.avgeriou@rug.nl}

\address[1]{School of Computer Science, Wuhan University, 430072 Wuhan, China}
\address[2]{Department of Mathematics and Computing Science, University of Groningen, 9747AG Groningen, The Netherlands}

\cortext[cor1]{Corresponding author}

\begin{abstract}
\noindent \textbf{Context}: As a software system evolves, its architecture tends to degrade, and gradually impedes software maintenance and evolution activities and negatively impacts the quality attributes of the system. The main root cause behind architecture erosion phenomenon derives from violation symptoms (i.e., various architecturally-relevant violations, such as violations of architecture pattern). Previous studies focus on detecting violations in software systems using architecture conformance checking approaches. However, code review comments are also rich sources that may contain extensive discussions regarding architecture violations, while there is a limited understanding of violation symptoms from the viewpoint of developers.\\
\textbf{Objective}: In this work, we investigated the characteristics of architecture violation symptoms in code review comments from the developers' perspective.\\
\textbf{Method}: We employed a set of keywords related to violation symptoms to collect 606 (out of 21,583) code review comments from four popular OSS projects in the OpenStack and Qt communities. We manually analyzed the collected 606 review comments to provide the categories and linguistic patterns of violation symptoms, as well as the reactions how developers addressed them.\\
\textbf{Results}: Our findings show that: (1) three main categories of violation symptoms are discussed by developers during the code review process; (2) The frequently-used terms of expressing violation symptoms are ``\textit{inconsistent}'' and ``\textit{violate}'', and the most common linguistic pattern is \textit{Problem Discovery}; (3) Refactoring and removing code are the major measures (90\%) to tackle violation symptoms, while a few violation symptoms were ignored by developers.\\
\textbf{Conclusions}: Our findings suggest that the investigation of violation symptoms can help researchers better understand the characteristics of architecture erosion and facilitate the development and maintenance activities, and developers should explicitly manage violation symptoms, not only for addressing the existing architecture violations but also preventing future violations.
\end{abstract}

\maketitle

\section{Introduction}\label{sec:Introduction}
During software evolution, the implemented architecture tends to increasingly diverge from the intended architecture. The resulting gap between the intended and implemented architectures is defined as \textit{architecture erosion} \cite{li2022SMS, Perry1992fssa}, and has been described using different terms in the literature and practice, such as architectural decay, degradation, and deterioration \cite{li2021uae, li2022SMS}. Architecture erosion can negatively affect quality attributes of software systems, such as maintainability, performance, and modularity \cite{li2021uae, Mendoza2021avd}.

Architecture erosion can manifest in a variety of symptoms during the software life cycle; such symptoms indicate that the implemented architecture is moving away from the intended one. In our recent systematic mapping study \cite{li2022SMS}, four categories of architecture erosion symptoms were reported: \textit{structural symptoms} (e.g., cyclic dependencies), \textit{violation symptoms} (e.g., violation of the layered pattern), \textit{quality symptoms} (e.g., high defect rate), and \textit{evolution symptoms} (e.g., rigidity and brittleness of the software system). From these four types of symptoms, violation symptoms are deemed as the most critical symptoms that practitioners should address because the accumulation of violation symptoms can render the architecture completely untenable \cite{DeSilva2012csa}. 
Violation symptoms of architecture erosion include various types of architecturally-relevant violations in software systems, such as violations of design principles, architecture patterns, decisions, requirements, modularity, etc. \cite{li2022SMS}. For the sake of brevity, we refer to violation symptoms of architecture erosion as \textit{violation symptoms} in the rest of this paper.

While a handful of temporary violation symptoms might be innocuous regarding the software system, the accumulation of architecture violations can lead to architecture erosion \cite{Perry1992fssa, DeSilva2012csa, Mendoza2021avd}, severely impacting run-time and design-time qualities. Therefore, identifying and monitoring violation symptoms is crucial to reveal inconsistencies between the implementation and the intended architecture; eventually this can help to, at least partially, repair architecture erosion \cite{li2022SMS, DeSilva2012csa}.

Prior studies focusing on erosion symptoms through inspecting source code might ignore implicit semantic information, while our work dives into violation symptoms by analyzing textual artifacts such as code review comments from the developers' perspective. In contrast to violation symptoms identification from source code using predefined abstract models \cite{Miranda2016acc, Pruijt2014srm} and rules \cite{Terra2009dcl, Rocha2017DCL, Caracciolo2015uaa, Juarez2017pee}, violation symptoms can also be detected by analyzing textual artifacts that contain information related to the system architecture and its design. Violation symptoms can occur at different stages of development and be self-admitted or pointed out by developers \cite{li2022sae}. In our previous study \cite{li2022sae}, we investigated two types of symptoms of architecture erosion in code reviews (i.e., structural and violations symptoms), and found that violation symptoms are the most frequently-discussed symptoms in architecturally-relevant issues during code review. Therefore, in this work, we focus on violation symptoms and attempt to categorize them and understand how developers express and address them.

Code review comments include rich textual information about the architecture changes that developers identified and discussed during development \cite{Paixao2019icr}. Code reviews are usually used to inspect defects in source code and help improve the code quality \cite{Bacchelli2013eoc}. 
Compared to pull requests that might not provide specific advice for development practice \cite{Li2021mpl}, code review comments provide a finer granularity of information for investigating architecture changes and violations from the developers' perspective.

Although valuable information on architecture violations is discussed during code review, there are no studies regarding the categories of violation symptoms that are discussed and admitted by developers, how developers express violation symptoms, and whether and how these symptoms are addressed during the development. To this end, we aim at understanding how developers discuss violation symptoms and providing an in-depth investigation to the categories of violation symptoms, as well as practical measures used to address them. We identified 606 (out of 21,583) code review comments related to architecture violations from four OSS projects in the OpenStack and Qt communities. The main contributions of this work are the following:
\begin{itemize}
    \item We created a dataset containing violation symptoms of architecture erosion from code review comments, which can be used by the research community for the study of architecture erosion.
    \item We identified the 606 violation symptoms and classified them into three categories with ten subcategories, as well as the ways that developers addressed these categories.
    \item This is the first study that investigated violation symptoms in textual artifacts (specifically, code review comments) from the perspective of practitioners.
    \item We identified the linguistic patterns of expressing architecture violation symptoms from code review comments.
\end{itemize}

The paper is organized as follows: Section~\ref{sec:Background} introduces the background of this study. Section~\ref{sec:Methodology} elaborates on the study design. The results of the research questions are presented in Section~\ref{sec:Results}, while their implications are further discussed in Section~\ref{sec:Discussion}. Section~\ref{sec:Threats} elaborates on the threats to validity. Section~\ref{sec:Related Work} reviews the research work of this study. Finally, Section~\ref{sec:Conclusions} summarizes this work and outlines the directions for future research.

\section{Background}\label{sec:Background}
In this section, we overview the background of our study regarding code review and architecture erosion with the corresponding erosion symptoms.

\subsection{Code Review}\label{sec:Code Review}
Code review is the process of analyzing assigned code for inspecting code and identifying defects. A methodical code review process can continuously improve the quality of software systems, share development knowledge, and prevent from releasing products with unstable and defective code. Currently, code review practices have become a crucial development activity that has been broadly adopted and converged to code review supported by tools. Moreover, tool-based code review has been widely used in both industry and open source communities. In recent years, many code review tools have been provided, such as Meta’s Phabricator\footnote{https://www.phacility.com/}, VMware’s Review-Board\footnote{https://www.reviewboard.org/}, and Gerrit\footnote{https://www.gerritcodereview.com/}.

Gerrit is a popular code review platform designed for code review workflows and is used in our selected projects (see Section~\ref{sec:Project Selection}). Once a developer submits new code changes (e.g., patches) and their description to Gerrit, the tool will create a page to record all the changes, and meanwhile the developer should write a message to describe the code changes, namely, a ``\textit{commit message}''. Gerrit conducts a sanity check to verify the patch is compliant and to make sure that the code has no obvious compilation errors. After the submitted patch passes the sanity check, code reviewers will manually examine the patch and provide their feedback to correct any potential errors, and then give a voting score. Note that, code reviewers cannot only comment on source code but also on code commits. The review and vote process will iterate with the purpose of improving the patch. Finally, the submitted patch will be merged into the code repository after passing the integration tests (i.e., without any issues and conflicts).

\subsection{Architecture Erosion}\label{sec:Architecture Erosion}
The sustainability of architecture depends on architectural design to ensure the long-term use, efficient maintenance, and appropriate evolution of architecture in a dynamically changing environment \cite{Venters2018ssr}. However, architecture erosion and drift are two essential phenomena threatening architecture sustainability. Architecture erosion happens due to the direct violations of the intended architecture, whereas architecture drift occurs due to extensive modifications that are not direct violations but introduce design decisions not included in the intended architecture \cite{Perry1992fssa, Venters2018ssr}.

The architecture erosion phenomenon has been extensively discussed in the past decades and has been described by various terms \cite{li2022SMS, li2021uae}, such as architecture decay \cite{Hassaine2012ADvISE, Le2018esa}, degradation \cite{Lenhard2019ess}, and degeneration \cite{Hochstein2005cad}. Architecture erosion manifests in a variety of symptoms during development and maintenance. A symptom is a (partial) sign or indicator of the emergence of architecture erosion. According to our recent systematic mapping study \cite{li2022SMS}, the erosion symptoms can be classified into four categories: \textit{structural symptoms} (e.g., cyclic dependencies), \textit{violation symptoms} (e.g., layering violation), \textit{quality symptoms} (e.g., high defect rate), and \textit{evolution symptoms} (e.g., rigidity and brittleness of systems). Previous studies have investigated different symptoms of architecture erosion. Mair \textit{et al}. \cite{Mair2014tfa} proposed a formalization method regarding the process of repairing eroded architecture through detecting violation symptoms and recommending optimal repair sequences. Le \textit{et al}. \cite{Le2018esa, Le2016rad} regarded architectural smells as structural symptoms and provided metrics to detect instances of architecture erosion by analyzing the detected smells. Bhattacharya \textit{et al}. \cite{Bhattacharya2007asm} developed a model for tracking software evolution by measuring the loss of functionality (as evolution symptoms). Regarding the scope of our work, we focus on the nature of architecture erosion (i.e., violation symptoms) through code review comments in this work, which paves the way towards shedding light on architecture violations from the developers' perspective.

\section{Methodology}\label{sec:Methodology}
The goal of this study is formulated by following the Goal-Question-Metric approach \cite{Basili1994gmq}: \textbf{analyze} \textit{code review comments} \textbf{for the purpose of} \textit{identification and analysis} \textbf{with respect to} \textit{violation symptoms of architecture erosion} \textbf{from the point of view of} \textit{software developers} \textbf{in the context of} \textit{open source software development}. 

\subsection{Research Questions}\label{subsec:RQs}
To achieve our goal, we define three Research Questions (RQs):

\begin{tcolorbox}\textbf{RQ1: What categories of violation symptoms do developers discuss?}\end{tcolorbox}
\noindent\textit{Rationale}: This RQ aims at investigating the categories of violation symptoms that frequently occur during the development process; an example of such a category is violations of architecture patterns. The proposed categories of violation symptoms in textual artifacts from code review comments can be used by practitioners as guidelines to avoid such violations in practice. For example, certain categories of violation symptoms may be associated with high erosion risks \cite{li2022SMS} and be regarded as important to provide warnings to developers. 

\begin{tcolorbox}\textbf{RQ2: How do developers express violation symptoms?}\end{tcolorbox}
\noindent\textit{Rationale}: Violation symptoms in code review comments are described in natural language, but there is a lack of evidence regarding how developers describe these violation symptoms. Specifically, we are interested in the terms and linguistic patterns\footnote{Grammatical rules that allow their users to speak properly in a common language \cite{DaSilvalpls2017}} that developers use to denote violation symptoms. Establishing a list of the terms and linguistic patterns used by practitioners can subsequently provide a basis for the automatic identification of violation symptoms through natural language processing techniques.

\begin{tcolorbox}\textbf{RQ3: What practices are used by developers to deal with violation symptoms?}\end{tcolorbox}
\noindent\textit{Rationale}: We aim at exploring what developers do when they encounter violation symptoms during the development process; this includes whether developers address the violation symptoms and how they do that. The answers to this RQ can help uncover best practices to cope with violation symptoms, and facilitate the development of methods and tools that promote such practices. 

\subsection{Project Selection}\label{sec:Project Selection}
To understand violation symptoms that developers face in practice, we selected four OSS projects from two communities, namely OpenStack and Qt; these projects have been commonly used in previous studies (e.g., \cite{Kashiwa2022ess}) due to their long development history and rich textual artifacts. OpenStack\footnote{https://www.openstack.org/} is a widely-used open source cloud software platform, on which many organizations (e.g., IBM and Cisco) collaboratively develop applications for cloud computing. Qt\footnote{https://www.qt.io/} is a toolkit and a cross-platform framework for developing GUIs, and is used by around one million developers to develop world-class products for desktop, embedded, and mobile operating systems.

Both OpenStack and Qt contain a large number of sub-projects, thus we selected two sub-projects from each community: Neutron and Nova from the OpenStack community, and Qt Base and Qt Creator from the Qt community (see Table~\ref{T:projects}). Neutron (providing networking as a service for interface devices) and Nova (a controller for providing cloud virtual servers) are mainly written in Python; Qt Base (offering the core UI functionality) and Qt Creator (the Qt IDE) are mainly developed in C++. The selected four projects are the most active projects in the OpenStack and Qt communities, respectively, and they are widely known for a plethora of code review data recorded in the Gerrit code review system \cite{Thongtanunam017rpm, Hirao2022crd}. 

\begin{figure*}[t]
	\centering
	\includegraphics[width=\textwidth]{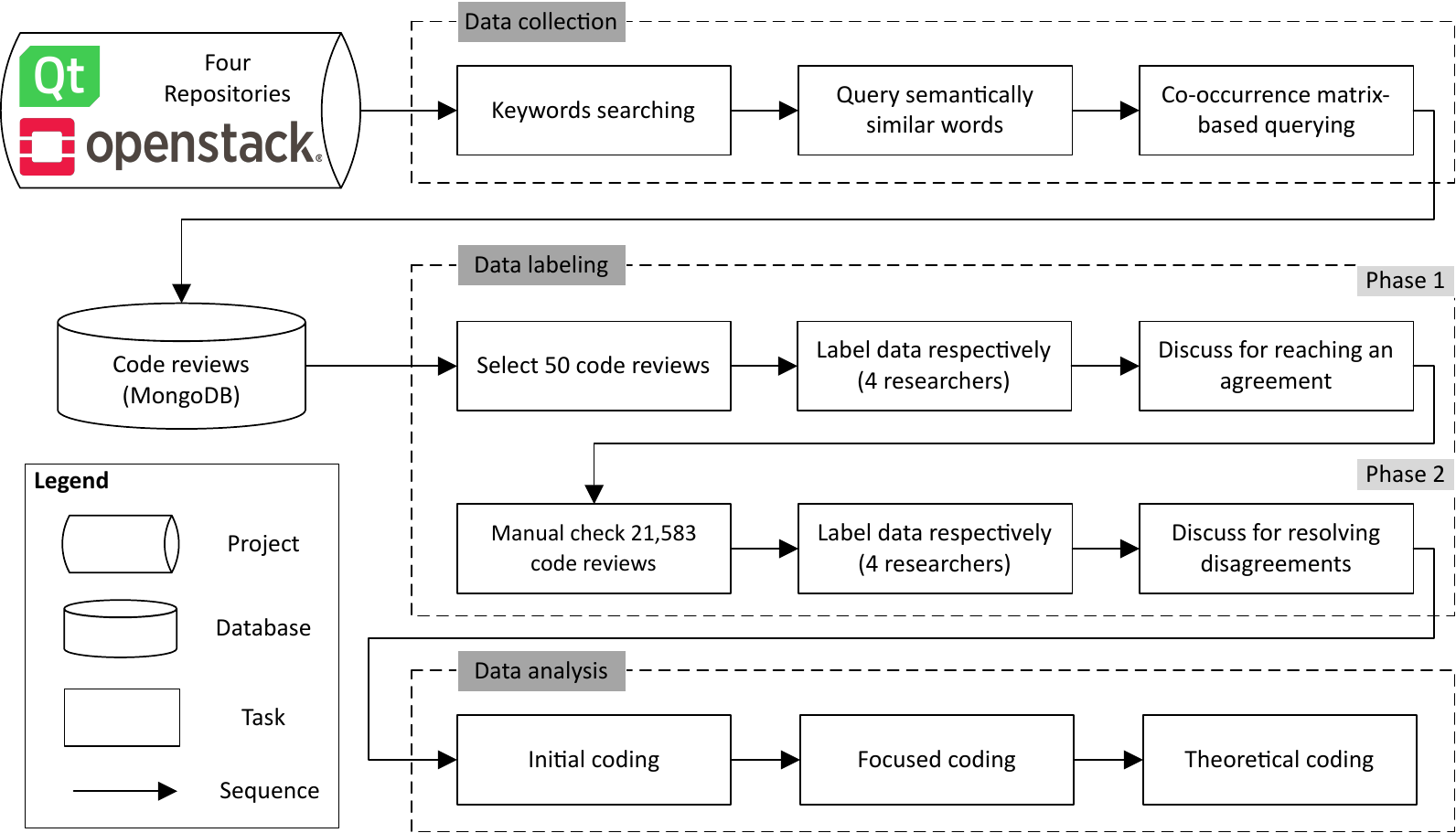}
    	\caption{An overview of the data collection and analysis process}\label{F:Overview}
\end{figure*}

\subsection{Data Collection}\label{sec:Data Collection}
An overview of our data collection, labelling, and analysis is shown in Figure~\ref{F:Overview}. Starting with the process of data collection (the top part of Figure~\ref{F:Overview}), we first employed Python scripts to mine code review comments (concerning source code and commits) of the four projects through the REST API\footnote{https://gerrit-review.googlesource.com/Documentation/rest-api.html} supported by the Gerrit tool. Then, we organized and stored the collected data in MongoDB. Our goal, as stated in the beginning of this section, is to analyze the violation symptoms that exist in code review comments from developers. Therefore, we removed the review comments that were generated by bots in the Qt community; we noticed that there were no code review comments generated by bots in the OpenStack community. However, manually analyzing the entire history of code review comments of the four projects is prohibitive in terms of both effort and time. Thus, we decided to collect the code reviews of the four projects in seven years between Jan 2014 and Dec 2020 to guarantee sufficient revisions for long-lived software systems. Finally, we obtained 518,743 code review comments concerning code and 48,113 review comments concerning commit messages from the four projects in the past seven years. Each item in our dataset contains review ID and patch information, including \texttt{change\_id}, \texttt{patch}, \texttt{file\_url}, \texttt{line}, and \texttt{message}; the message variable includes code review comments concerning source code and commits. All the scripts and the dataset of this work have been made available in the replication package \cite{Replication}.

\begin{table*}[htb]
    \centering
    \caption{An overview of the selected projects}\label{T:projects}
    \begin{tabular}{|p{15mm}|p{30mm}|p{50mm}|m{12mm}<{\centering}|m{20mm}<{\centering}|m{20mm}<{\centering}|}
    \hline
    \textbf{Project} & \textbf{Domain} & \textbf{Repository} & \textbf{Language} & \textbf{\#Review comments of code} & \textbf{\#Review comments of commits}\\\hline
    Nova & Virtual server management & \url{https://opendev.org/openstack/nova} & Python & 152,107 & 15,164\\\hline
    Neutron & Network connectivity & \url{https://opendev.org/openstack/neutron} & Python & 181,839 & 16,719\\\hline
    Qt Base & Providing UI functionality & \url{https://code.qt.io/cgit/qt/qtbase.git/} & C++ & 123,546 & 13,369\\\hline
    Qt Creator & A cross-platform IDE & \url{https://code.qt.io/cgit/qt-creator/qt-creator.git} & C++ & 61,251  & 2,861\\\hline
    \multicolumn{4}{|c|}{\textbf{Total}} & \textbf{518,743} & \textbf{48,113}\\\hline
    \end{tabular}
\end{table*}

The collected review comments contain a large number of entries, such as ``\textit{Done}'' and ``\textit{Ditto}'', which are not related to the discussion on violation symptoms. To effectively collect and locate the associated code review comments on violation symptoms of architecture erosion, we decided to employ a keyword-based search approach. We employed the keywords presented in our previous work \cite{li2022sae} (see the coarse-grained keywords in Table~\ref{T:keywords}) and improved the keyword set (see the fine-grained keywords in Table~\ref{T:keywords}) as described below. Specifically, to derive possible and associated synonyms of the keywords in software engineering practices, we adopted the pre-trained word2vec model proposed by Efstathiou \textit{et al}. \cite{Efstathiou2018weso} for querying semantically similar terms. The authors of \cite{Efstathiou2018weso} trained this model with over 15GB of textual data from Stack Overflow posts, which contain a plethora of textual expressions and words in the software engineering domain. We utilized this pre-trained word embedding model to query similar terms of the coarse-grained keyword set. For example, we got ``\textit{discrepancy}'' and ``\textit{deviation}'' which are similar terms to ``\textit{divergence}'', and then the first two authors discussed together to manually check and remove unrelated and duplicate words, such as ``\textit{oo}'' which is related to programming languages rather than architecture violations. The keywords set used to search code review comments includes both the coarse-grained keywords and the fine-grained keywords listed in Table~\ref{T:keywords}.

In addition, given that the effectiveness of the keyword-based approach highly depends on the set of keywords, we chose the iterative approach proposed by Bosu \textit{et al}. \cite{Bosu2014icc} to further improve the keyword set by adding possible keywords that are related to the keywords in Table~\ref{T:keywords}. This approach has already been employed in previous studies (e.g., \cite{li2022sae, Han2021ucs}) that used keyword-based search in code review data. We implemented this approach in the following steps:
\begin{enumerate}
    \item Search in the collected review comments using the keyword set in Table~\ref{T:keywords}, and then establish a corpus by collecting the relevant comments that encompass at least one keyword from the keyword set (e.g., ``\textit{violation}'').
    \item Process the retrieved code review comments that contain at least one keyword in our keyword set and remove English stopwords, punctuation, code snippets, and numbers. That results in a list of tokens.
    \item Conduct a stemming process (using SnowballStemmer from the NLTK toolkit \cite{Bird2010nlp}) to obtain the stem of each token (e.g., “\textit{architecture}” and “\textit{architectural}” have the same token “\textit{architectur}”).
    \item Build a document-term matrix from the corpus, and find additional words that co-occur frequently with each of our keywords (co-occurrence probability of 0.05 in the same document).
    \item Manually check and discuss the list of frequently co-occurring additional words to determine whether the newly discovered words should be added to the keyword set.
\end{enumerate}

After executing this approach, we have not found any keywords that co-occur with the keywords based on a co-occurrence probability of 0.05 in the same document. Therefore, we believe that we have minimized the possibility of missing potentially co-occurred and associated words, and the keyword set could be relatively adequate and comprehensive for the search in this study. The keywords used in this work are presented in Table~\ref{T:keywords}. In total, we collected 21,583 code review comments from the four OSS projects that contain at least one keyword.

\begin{table}[htb]
    \centering
    \caption{Keywords related to violation symptoms of architecture erosion}\label{T:keywords}
    \begin{tabular}{p{80mm}}
    \hline
    \cellcolor{gray}\textbf{Coarse-grained keywords}\\\hline
    architecture, architectural, structure, structural, layer, design, violate, violation, deviate, deviation, inconsistency, inconsistent, consistent, mismatch, diverge, divergence, divergent, deviate, deviation\\\hline
    \cellcolor{gray}\textbf{Fine-grained keywords}\\\hline
    layering, layered, designed, violates, violating, violated, diverges, designing, diverged, diverging, deviates, deviated, deviating, inconsistencies, non-consistent, discrepancy, deviations, modular, module, modularity, encapsulation, encapsulate, encapsulating, encapsulated, intend, intends, intended, intent, intents, implemented, implement, implementation, as-planned, as-implemented, blueprint, blueprints, mis-match, mismatched, mismatches, mismatching\\\hline
    \end{tabular}
\end{table}


\subsection{Data Labeling and Analysis}\label{sec:Data Analysis}

We filtered out a large number of irrelevant code review comments in Section~\ref{sec:Data Collection}. Still, the retrieved code review comments that contain at least one keyword might be unrelated to violation symptoms. Thus, we needed to manually check and further remove these semantically unrelated review comments. We conducted \textbf{data labeling} in two phases, as illustrated in Figure~\ref{F:Overview}.

\textbf{Phase 1}. We decided to conduct a pilot data labeling to reach a consensus and to ensure that we have the same understanding of violation symptoms. Four researchers (the first author and three master students) had an online meeting to discuss the characteristics of violation symptoms. We randomly selected 50 review comments from the collected data. The four researchers independently labeled the violation symptoms from the review comments via MS Excel sheets, and provided reasons for their labeling results. Then the four researchers had another meeting to check the similarities and differences between their labeling results for reaching an agreement, and any disagreements were discussed with the second author to reach a consensus. Note that, during the data labeling process, the researchers not only read the text content of the code review comments per se, but also read their corresponding code snippets, documentation, and commit messages. This helped us further mitigate the threat of wrong labels, such as simple violations at the code level (e.g., pep8 coding style violation\footnote{https://peps.python.org/pep-0008/}). In the end, to measure the inter-rater agreement between the researchers, we calculated the Cohen's Kappa coefficient value \cite{Cohen1960aca} of the pilot data labeling and got an agreement value of 0.857, which demonstrates a substantial agreement between them.

\textbf{Phase 2}. After the pilot data labeling, the four researchers started the formal data labeling by dividing the retrieved 21,583 code review comments into four parts; each researcher manually labeled one fourth of this dataset (almost 5,400 review comments). The first author created the MS Excel sheets and shared them with the other three researchers. The four researchers were asked to label the textual information associated with violation symptoms. After the formal data labeling, the first author checked the data labeling results from the other three researchers to make sure that there were no false positive labeling results. To mitigate potential bias, we discussed all the conflicts in the labeling results until we reached an agreement. In other words, the data labeling results were checked by at least two researchers. 
The researchers followed the same process as in Phase 1 to conduct data labeling.

Finally, for \textbf{data analysis}, we employed Constant Comparison \cite{Charmaz2014CC, Stol2016gt} to analyze and categorize the identified textual information. Constant Comparison \cite{Charmaz2014CC, Stol2016gt} can be used to yield concepts, categories, and theories through a systematic analysis of qualitative data. The Constant Comparison process according to Charmaz \textit{et al}. \cite{Charmaz2014CC} includes three steps. The first step is \textit{initial coding} executed by the four researchers, who examined the review comments by identifying violation symptoms from the retrieved textual information. Second, we applied \textit{focused coding} executed by the first author and reviewed by the second author, by selecting categories from the most frequent codes and using them to categorize the data. For example, ``\textit{feels like this DB work violates the level of abstraction we are expecting here}'' was initially coded as \textit{violation of abstraction}, and we merged this code into \textit{violation of design principles}, as we considered that the violation of ``\textit{the expected level of abstraction}'' belongs to \textit{violation of design principles}. Third, we applied \textit{theoretical coding} to specify the relationship between codes. We checked the disagreements on the coding results by the four researchers to reduce the personal bias, and discussed the disagreements with the second author to get a consensus. The whole manual labeling and analysis process took the researchers around one and a half months. 

During the data analysis process, if the violation symptoms were specifically stated, we assigned them to specific groups. Conversely, when the symptoms were defined more broadly or lacked specificity, we classified them into general categories. We relied solely on the explicit textual content of the comments themselves during data analysis, without subjective interpretation. Besides, we note that we did not find multiple violation symptoms were discussed in a single review comment. In other words, each identified review comment has one single label. In addition, unlike our previous study \cite{li2022sae} which focused on both structural and violation symptoms, we conducted the data collection and labeling processes in this study from scratch by following the aforementioned steps; consequently we have established a more comprehensive and larger dataset on violation symptoms \cite{Replication}.



\section{Results}\label{sec:Results}

\subsection{Overview}\label{sec:Overview}

Before delving into the findings of the three RQs, we briefly report an overview of the descriptive statistics about the identified violation symptoms from the selected four projects.

\begin{figure}[t]
	\centering
	\includegraphics[width=0.48\textwidth]{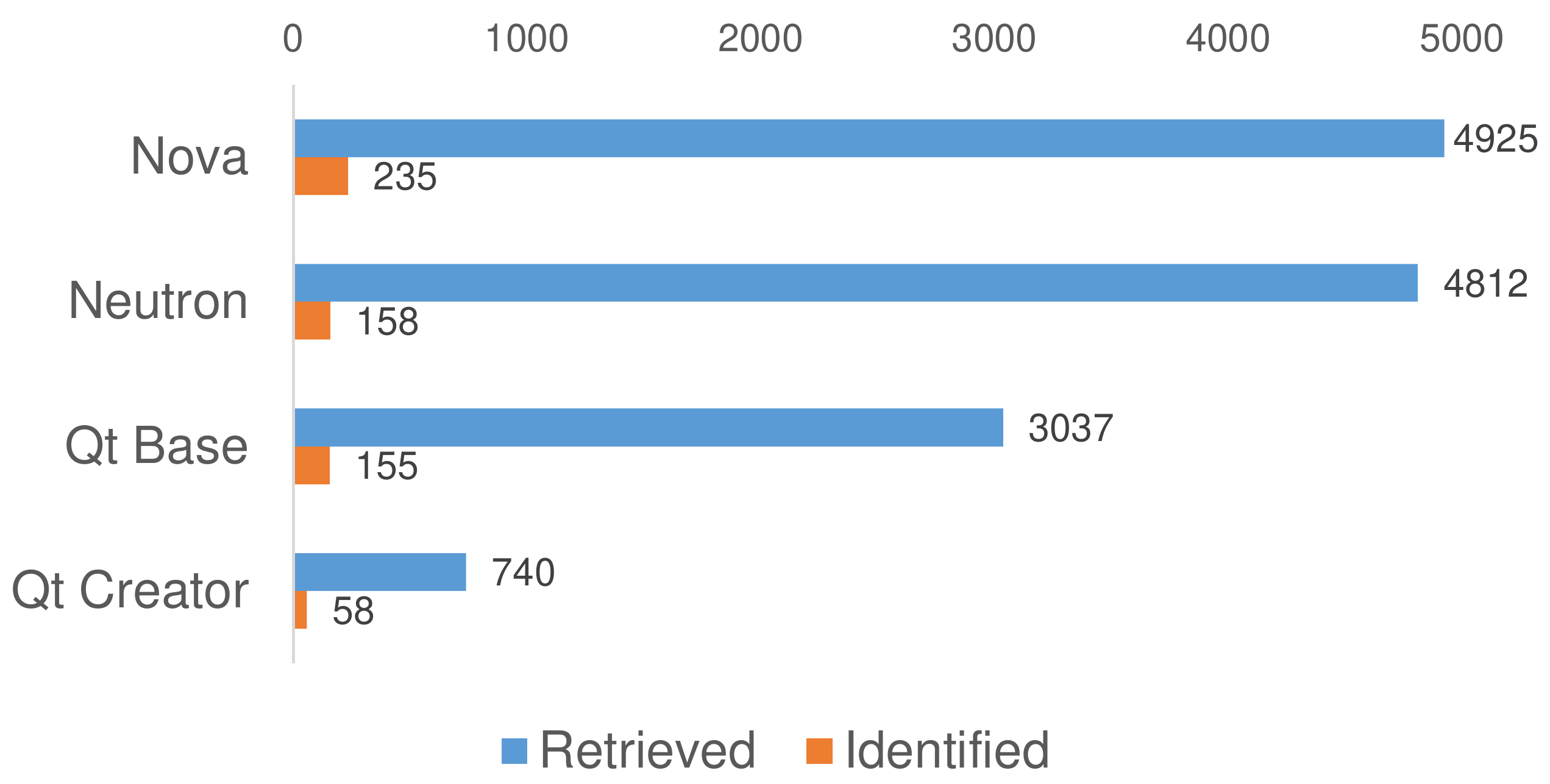}
    	\caption{Overview of the retrieved review comments containing the violation symptom keywords and the identified review comments related to violation symptoms in the four projects}\label{F:overview_dataset}
\end{figure}

Figure~\ref{F:overview_dataset} shows the retrieved review comments containing the violation symptom keywords in Table~\ref{T:keywords} and the identified review comments related to violation symptoms from the four projects. We observed that (1) the proportion of retrieved review comments across the four projects aligns closely with the results presented in Table~\ref{T:projects}. Specifically, the percentages are as follows: Nova at 2.94\%, Neutron at 2.42\%, Qt Base at 2.22\%, and Qt Creator at 1.15\%; (2) the identified review comments account for a similar percentage of the retrieved review comments across the four projects; the respective percentages are: Nova at 4.77\%, Neutron at 3.28\%, Qt Base at 5.10\%, and Qt Creator at 7.84\%.

In terms of the identified code review comments related to violation symptoms in our dataset, only a small portion of these comments (59 out of 606, 9.7\%) pertained to the content in commit messages. In contrast, the vast majority of the identified review comments related to violation symptoms (547 out of 606, 90.3\%) were associated with source code. Considering that the number of review comments on source code is 10 times higher than the number of review comments on commit messages as shown in Table~\ref{T:projects}, the proportion (10:1) is similar to the proportion of the identified review comments related to violation symptoms from the two sources (547:59).

\subsection{RQ1 - Categories of Violations Symptoms}\label{sec:Results_RQ1}
To answer RQ1, we identified 606 (out of 21,583) code review comments from the four projects that contain a discussion of violation symptoms of architecture erosion. As shown in Table~\ref{T:Categories}, the collected code review comments can be classified into three categories of violation symptoms, with ten subcategories as follows.
\begin{itemize}
    \item \textcolor{blue}Design-related violation: six types of violation symptoms pertained to design are identified, including: structural inconsistencies, violation of design decisions, violation of design principles, violation of rules, violation of architecture patterns, and violation of database design.
    \item Specification-related violation: two types of violation symptoms related to specifications are identified, including: violation of documentation and violation of API specification.
    \item Requirement-related violation: two types of violation symptoms relevant to requirements are identified, including: violation of architecture requirements and violation of constraints.
\end{itemize}

In the following  subsections, we present the detailed descriptions of the ten subcategories, accompanied by a range of real-world examples. The ten subcategories of violation symptoms are presented according to their frequencies in Table~\ref{T:Categories} within the dataset.

\begin{table*}[htb]
    \centering
    \caption{Categories of violation symptoms in code review comments}\label{T:Categories}
    \begin{tabular}{|p{22mm}|p{40mm}|p{85mm}|m{9mm}<{\centering}|}\hline
    \textbf{Category} & \textbf{Subcategory} & \textbf{Description} & \textbf{Count} \\\hline
    Design-related violation & Structural inconsistencies & Violations of consistencies of structural design in architecture (e.g., architectural mismatch) that exist in various architectural elements (e.g., components, ports, modules, and interfaces). & 205 \\\cline{2-4}
    ~ & Violation of design decisions & Violations of selected design decisions, including design rationale, intents, or goals that may cause implementation errors and give rise to ever-increasing maintenance costs. & 92 \\\cline{2-4}
    ~ & Violation of design principles & Violations of the common design principles (e.g., the SOLID principles) or divergences from object-oriented development guidelines. & 91 \\\cline{2-4}
    ~ & Violation of rules & Violations of the predefined architecture rules or policies when the implementation does not actually follow the rules. & 46 \\\cline{2-4}
    ~ & Violation of architecture patterns & Violations of architecture patterns (e.g., violations of layered pattern) when the architecture pattern implementations do not conform to their definitions. & 33\\\cline{2-4}
    ~ & Violation of database design & Violations of the principles or constraints in designing the database of systems. & 25 \\\hline

    Specification-related violation & Violation of documentation & Violations of the specification in development documents that hinder architecture improvements and modifications. & 56 \\\cline{2-4}
    ~ & Violation of API specification & Violations or inconsistencies of the API claims or specification. & 36\\\hline
    
    Requirement-related violation & Violation of architecture requirements & Violations of the intended requirements (e.g., user requirements) during the development and maintenance process. & 11 \\\cline{2-4}
    ~ & Violation of constraints & Violations of specific constraints imposed by the intended architecture may have an impact on architecture design. & 11 \\\hline   
    \end{tabular}
\end{table*}

\subsubsection{\textbf{Structural inconsistencies}} 
Structural inconsistencies occur due to various reasons, for example, it might be associated with a change of software specifications, and related components that implement the specifications cannot be automatically updated to keep consistency \cite{Grundy1998imm}. 
Structural inconsistencies can delay system development, increase the cost of development process, and further jeopardise the properties of system quality (e.g., reliability and compatibility). For example, one developer mentioned how inconsistency leads to an increased number of collaborating modules (i.e., module dependencies), and consequently the complexity of the system: 

\noindent\faComment ``\textit{Consistency is a nice thing, especially as in this case by not having consistency we are increasing the number of collaborating modules this module has.}''

%
Another example is related to inconsistent implementation between extension classes:

\noindent\faComment ``\textit{I think it could, but that would make it inconsistent with every other extension class's implementation of this method (I gripped). I think its best to keep them consistent. Feel free to file a low-hanging-fruit bug to fix this across all extensions.}''


In addition, architectural mismatch is another typical inconsistency issue in this category, which denotes the inability to successfully integrate component-based systems \cite{Garlan1995amw}. Architectural mismatch happens when there are conflicting assumptions among architecture elements (e.g., connectors and components)  during development \cite{Garlan1995amw, Garlan2009archm}. One reviewer mentioned an architectural mismatch between the server and agent side due to port design issues:

\noindent\faComment ``\textit{The port security extension adds functionality to disable port security, which is on by default. I don't think we should be changing the default behavior when port security is not present. User can explicitly disable port security if needed with port security extension. Enabling it here also makes a mismatch between server and agent side code.}''

\subsubsection{\textbf{Violation of design decisions}} This category contains violations of design decisions that have been made, especially the violation of design rationale. Such violations may lead to implementation errors and consequently aggravate maintenance costs. For example, one developer stated:

\noindent\faComment ``\textit{I chose olso because the intent is for this file to eventually not use anything in nova, so adding a nova import seemed like a step in the wrong direction.}''

In addition, another reviewer pointed out a violation of design intent:

\noindent\faComment ``\textit{But if \_get\_provider\_traits is being updated in this otherwise aggregate-specific change, it seems like we're splitting in two different directions which defeats the purpose of trying to be consistent.}''

Such violations imperil the actual implementation and lead to high error-proneness, and make software systems harder to implement, comprehend, maintain, and evolve. Here is an implementation error due to violating certain design decision as one developer mentioned:

\noindent\faComment ``\textit{Originally, QMT\_CHECK was Q\_ASSERT (before I added this code to QtCreator). It is a serious implementation error ... If you are not happy with a crash you can add a check against 0. This will avoid the crash here but I am pretty sure that it will crash sooner or later on a different location.}''


\subsubsection{\textbf{Violation of design principles}} The identified violation symptoms in this category denote violations of common design principles in object-oriented design and development, such as encapsulation, abstraction, and hierarchy \cite{Martin2006app}. Common examples of object-oriented design principles include the SOLID principles proposed by Martin \cite{Martin2003asd}, i.e., \textit{Single Responsibility Principle}, \textit{Open Closed Principle}, \textit{Liscov Substitution Principle}, \textit{Interface Segregation Principle}, and \textit{Dependency Inversion Principle}. As an example, one reviewer pointed out a violation of the Interface Segregation Principle:

\noindent\faComment ``\textit{But here don't we have to make a upcall from compute to api db, which will violate api/cell isolation rules. Is there any workaround in this case?}''

Another two examples are violations of abstraction and encapsulation:

\noindent\faComment ``\textit{Having said all of that: I get that I'm violating an abstraction layer in LinkLocalAddressPair and that this is surprising (and therefore bad).}''

\noindent\faComment ``\textit{It would seem to violate encapsulation to have to know to set the default value for an attribute outside of the object.}''

\subsubsection{\textbf{Violation of documentation}} The identified violation symptoms in this category encompass violations of instructions in documentation, e.g., not following the instructions on how to implement an interface. Such violations of documentation can hinder the subsequent architecture improvements and modifications \cite{Macia2012codeanomalies}. Two examples regarding violations of documentation are presented below:

\noindent\faComment ``\textit{The case of physical or service VM routers managed by L3 plugins doesn't appear to be supported here when the gateway IP is not or cannot be set to the router LLA. By supporting those cases in this way though, we're breaking the reference implementation as documented.}''

\noindent\faComment ``\textit{The reimplementation in QStandardItemModel does *not* match the documentation as it replaces the entire set of roles with the new ones. It also violates the documentation with the silly EditRole->DisplayRole thing (touching the value for a role not passed in input; although one could say that QSIM `aliases' EditRole with the DisplayRole in all cases).}''



\subsubsection{\textbf{Violation of rules}} The violations of architecture rules occur when the implementation does not actually follow the predefined rules by architects. Different systems have their own defined architecture rules or policies. During the development, rules provide the way to specify the allowed and specific relationships between architecture elements (e.g., components and modules). 
As mentioned by a developer:

\noindent\faComment ``\textit{... this implies that that caller is able to put constraints on the driver which may violate the rules built into the driver.}''

Another violation of predefined rules pointed out by a reviewer is:

\noindent\faComment ``\textit{We would now be limiting new spawns to only be allowed to the host that was the cause of the violation, thus causing the violation to be made worse. But maybe this is okay, since there isn't much we can do if the policy has been violated prior to this.}''

\subsubsection{\textbf{Violation of API specification}} All API-related violations and inconsistencies belong to this category. API documentation describes the explicit specification of interfaces and dependencies. Due to the changing business requirements and continuous demands to upgrade functionalities, API evolution is inevitable. Violations of API specification encompass improper API usages and inconsistent API calls. Improper applications of APIs can give rise to unexpected system behaviors and eventually cause architectural inconsistencies. For example, developers do not adhere to the contract or specification to use the required APIs. As one reviewer stated:

\noindent\faComment ``\textit{the point is that with neither a category backend nor completely disabled output, the macro should be unusable (even if it would print something, the category would be missing, i.e., the implementation would violate the api).}''

Besides, inconsistent API calls can also cause inconsistencies, such as getting different responses from different versions of APIs (e.g., distinct parameters and returns). As one developer mentioned:

\noindent\faComment ``\textit{On v2 API, there is a lot of API inconsistencies and we need to fix them on v3 API. So we can change API parameters on v3 API.}''

\subsubsection{\textbf{Violation of architecture patterns}} Architecture patterns provide general and reusable solutions for particular problems, such as the layered pattern and client-server pattern. Violating architectural patterns undermines the sustainability and reliability of software systems and increases the risk of architecture erosion. For example, we found that violations of the layered pattern are one of the most common types of this category. Modern software systems often contain millions of lines of code across many modules. Therefore, employing hierarchical layers is a common practice to organize the relationships between modules. Violations of layered systems can negatively impact the quality attributes (e.g., reusability, maintainability, and portability), eventually leading to architecture erosion. For instance, a developer commented on a violation of the layered pattern:

\noindent\faComment ``\textit{That all said, I'm definitely -2 (even if not core ;-) ) on that patch, because I think it's a layer isolation violation to just make the call here. It should be fixed at the Compute API level rather IMHO.}''

Another example regarding the layered pattern violation is:

\noindent\faComment ``\textit{We could get some race conditions when starting the scheduler where it would not know the allocation ratios and would have to call the computes, which is a layer isolation violation to me.}''

\subsubsection{\textbf{Violation of database design}} 
Databases are one of the key architectural elements \cite{Bass2021sap}, and can negatively impact the system quality attributes when their design is violated. This category includes the problems caused by code changes that violate the constraints of database design, such as primary and foreign key constraints. For example, as mentioned by one reviewer, the network port will no longer exist when a foreign key violation is generated:

\noindent\faComment ``\textit{If subnet was fetched in reference IPAM driver, port got deleted and foreign key violation was generated on ipallocation insert (because port no longer exists).}''

Another example is about unique key violation:

\noindent\faComment ``\textit{In the bug report, the randomly generated index happens to be 2, which violates the already existing (router, 1) unique key constraints.}'' 


\subsubsection{\textbf{Violation of constraints}} Constraints are pre-determined special design decisions with zero degrees of freedom \cite{Bass2012sap}, which can be regarded as special requirements that cannot be negotiable and impact certain aspects of architecture implementation. Such violations often denote the concrete statements, expressions, and declarations in source code that do not comply with the constraints imposed by the intended architecture \cite{Terra2015rsr}, such as inter-module communication constraints. For example, one reviewer mentioned a constraint violation in the system:

\noindent\faComment ``\textit{If a specific subnet is passed in, then the IPAM system would try to give that subnet, but if it's already use or it violates the constraints in the IPAM system, it would refuse.}''

Another example concerning constraint violation is:

\noindent\faComment ``\textit{This should probably be HTTPBadRequest: the provided allocation has a form that violates Inventory constraints, so if the allocation (the request body) changes, it could work.}''


\subsubsection{\textbf{Violation of architecture requirements}} 
Requirements and especially quality attribute requirements are closely related to the architecture of a system. Architecturally significant requirements drive the architecture \cite{Bass2021sap}, but are, unfortunately, commonly violated \cite{li2022SMS}. Moreover, architecturally significant requirements specify the major features and functionalities that a particular product should include, and convey the expectations of the stakeholders for the software product. As an example, a developer mentioned that the existing code had violated the requirements:

\noindent\faComment ``\textit{We don't have this in our requirements ... I guess this was already violated by existing code so I don't really want to block on it, we can handle it separately.}''

Another example of violation of requirements is:

\noindent\faComment ``\textit{This is global/module level data so it will get loaded once and stay loaded for the life time of the wsgi process even if the application is restrated in the interpreter. we are the ortinial code violated the requirement in pont3 of ...}''


\subsection{RQ2 - Expression of Violation Symptoms}\label{sec:Results_RQ2}
\begin{figure}[htb]
	\centering
	\includegraphics[width=0.48\textwidth]{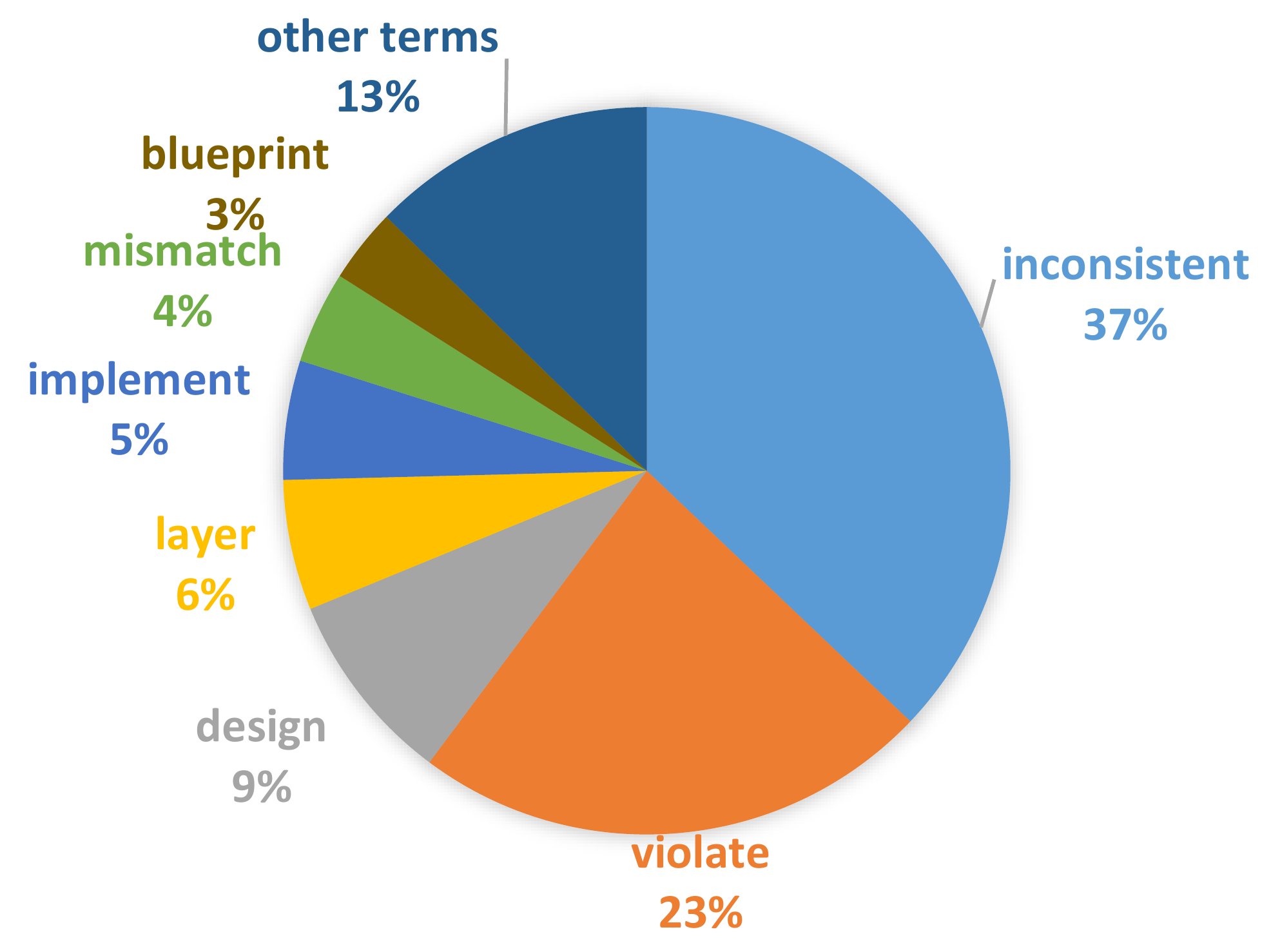}
    	\caption{Distribution of the frequently used terms related to description of violation symptoms}\label{F:terms}
\end{figure}

For answering RQ2, we inspected the content of the identified violation symptoms in code review comments to identify the frequently-used terms and associated linguistic patterns. Figure~\ref{F:terms} presents the distribution of the most frequently-used terms related to the discussion of violation symptoms in review comments. We use typical terms to represent the words which have the same meaning. For example, ``\textit{inconsistent}'' contains all the terms that have the same meaning, such as \textit{not consistent}, \textit{inconsistent}, \textit{inconsistency}, and \textit{inconsistencies}. The most frequently-used term is ``\textit{inconsistent}'' (37\%, 225 out of 606) and is related to the notion of ``\textit{consistency}''. ``\textit{Violate}'' comes second with 140 (23\%, out of 606) comments, followed by ``\textit{design}'' (9\%, 52 out of 606) and ``\textit{layer}'' (6\%, 35 out of 606). We put the less-frequently used terms into ``\textit{other terms}'', such as ``\textit{module}'' and ``\textit{architecture}''.

\begin{table*}[t]
    \centering
    \caption{Categories and percentages of linguistic patterns used to express violation symptoms in code review comments}\label{T:Categories and Percentages}
    \begin{tabular}{|l|p{39mm}|p{75mm}|m{15mm}<{\centering}|}\hline
    \textbf{Linguistic Patterns} & \textbf{Description} & \textbf{Example} & \textbf{Percentage}\\\hline
    Problem Discovery & Linguistic patterns related to unexpected or unintended behaviors & ``\textit{Well, so this patch actually makes the API worse: It is as unclear as before when a temporary file is created in the temporary directory and when not, *and* it deviates from the behavior of QTemporaryFile as well, so anyone knowing that behavior will get unexpected results here.}'' & 90.4\%  \\\hline
    Solution Proposal & Linguistic patterns related to describe possible solutions for founded problems & ``\textit{I'm not convinced that it is wise to deviate from that with this one, other than to use QPlainTestLogger::outputMessage(), which has side-effects on WinCE, Windows and Android. ... Please consider inheriting QAbstractTestLogger instead and calling QAbstractTestLogger::outputString() instead of QPlainTestLogger::outputMessage().}'' & 10.4\% \\\hline
    Opinion Asking & Linguistic patterns used for inquiring someone about his/her viewpoints and thoughts & ``\textit{Why diverge from the pattern established for toFoo\_helper below?}'' & 6.8\% \\\hline
    Information Giving & Linguistic patterns for informing someone about something & ``\textit{I didn't want to optimize in this way, because the code is wrong: The parent implementation should only be called if the sub-class does not handle the line itself. Several implementations got this wrong, and it is indeed not self-explanatory. I will fix all of those as part of moving away from the chaining approach.}'' & 4.8\% \\\hline
    Feature Request & Linguistic patterns for providing suggestions/recommendations/ideas & ``\textit{Recently there were some changes in API, virtual getters were replaced with protected setter. Please be consistent with it and provide protected setter and public getter, both non-virtual. ... I saw those changes came from them. And I strongly agree with it.}'' & 2.0\%\\\hline
    Information Seeking & Linguistic patterns related to ask help or information from others & ``\textit{I'm a bit confused as to how this violates the open/closed principle. ... I'm a bit unclear as to what your are suggestion as an alternative.  Are you suggesting having a separate config option for each traffic type as an alternative?}'' & 1.8\%\\\hline
    \end{tabular}
\end{table*}

In addition, to further analyze the characteristics of the comments related to violation symptoms, we summarized and categorized the linguistic patterns of expressing violation symptoms. We discovered the linguistic patterns by reviewing words or phrases that either frequently appear or are relatively unique to a particular category. Subsequently, we manually checked and categorized the comments into six categories based on the linguistic patterns proposed by Di Sorbo \textit{et al}. \cite{di2016deca}, namely, \textit{Feature Request}, \textit{Opinion Asking}, \textit{Problem Discovery}, \textit{Solution Proposal}, \textit{Information Seeking}, and \textit{Information Giving}, which have been employed to identify the linguistic patterns used in various textual artifacts (e.g., app reviews \cite{di2016wwu, di2017ufb} and issue reports \cite{Huang2018aim}). Table~\ref{T:Categories and Percentages} presents the statistical results of the review comments related to violation symptoms, including the categories, descriptions, examples, and percentages. A few code review comments contain more than one linguistic pattern, and overall they add up to more than 100\%. Our results show that most (90.4\%) of the comments related to violation symptoms are about \textit{Problem Discovery}, followed by \textit{Solution Proposal} (10.4\%) and \textit{Opinion Asking} (6.8\%). Moreover, we list the linguistic patterns (frequency $\geq$ 3) used to express violation symptoms in Table~\ref{T:linguistic patterns}.

\begin{table}[htb]
    \centering
    \caption{Linguistic patterns (frequency $\geq$ 3) used to express violation symptoms in code review comments}\label{T:linguistic patterns}
    \begin{tabular}{|m{5mm}<{\centering}|p{72mm}|}\hline
    \textbf{\#} & \textbf{Problem Discovery} \\\hline
    1 & This is breaking [something]\\\hline
    2 & This seems to violate [something]\\\hline
    3 & It seems like a (layer) violation of [something]\\\hline
    4 & It seems inconsistent/not consistent with [something]\\\hline
    5 & [someone] probably have [an issue]\\\hline
    6 & This violates/breaks [something]\\\hline
    7 & [someone] is/are violating the rules\\\hline
    8 & This is a violation of [something]\\\hline
    9 & This is not consistent with [something]\\\hline
    10 & [something] is/are inconsistent with [something]\\\hline
    11 & There will be inconsistencies in [something]\\\hline
    12 & There are inconsistencies [between something]/[of/in something]\\\hline
    13 & [something] leads to inconsistency \\\hline
    14 & [something] is a (poor/awful/terrible) design mistake/flaw/choice \\\hline
    15 & [something] diverges/deviates from [something]\\\hline
    \textbf{\#} & \textbf{Solution Proposal} \\\hline
    1 & I think [someone] should/need to [verb + subject]  \\\hline
    2 & [someone] should modify/preserve/revise [something] \\\hline
    3 & We should remove [something]\\\hline
    4 & I think [do something] would better if [something]\\\hline
    5 & [someone] should/need to stay/keep consistent with [something]  \\\hline
    6 & To fix it, we need to [do something] \\\hline
    7 & I think what we should do is [something] \\\hline
    8 & Did you consider the approach of [something] \\\hline
    
    \textbf{\#} & \textbf{Opinion Asking} \\\hline
    1 & Why are you diverging from [something]?\\\hline
    2 & Would it be possible to [do something]?\\\hline
    3 & Maybe we should [do something], what do you think about it?\\\hline
    4 & Should/would we [do something]?\\\hline
    
    \textbf{\#} & \textbf{Feature Request} \\\hline
    1 & We should/need to keep consistent with [something] \\\hline
    2 & It is better to [do something]\\\hline
    3 & I wonder if we can [do something]\\\hline
    
    \textbf{\#} & \textbf{Information Giving} \\\hline
    1 & I will modify/fix [something] to address/keep consistent with [something]\\\hline
    2 & That is why I [doing something]\\\hline
    
    \textbf{\#} & \textbf{Information Seeking} \\\hline
    1 & Is there a different [something]? \\\hline
    2 & Are you suggesting [something]?\\\hline
    3 & Are we planning to [do something]?\\\hline
\end{tabular}
\end{table}

\subsection{RQ3 - Dealing with Violation Symptoms}\label{sec:Results_RQ3}
\begin{figure}[htb]
	\centering
	\includegraphics[width=0.49\textwidth]{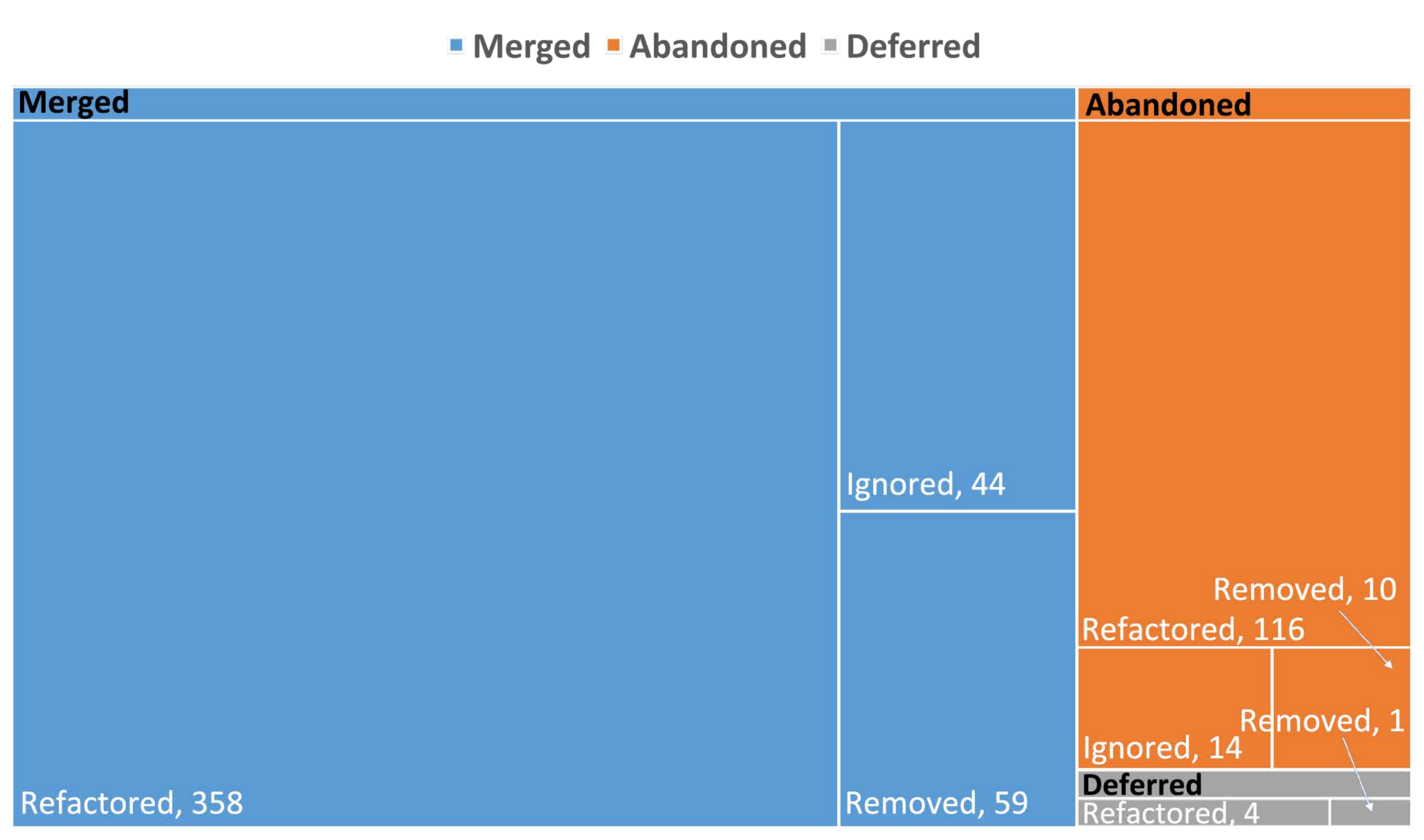}
    	\caption{Distribution of the developers' reactions in response to violation symptoms in code review comments}\label{F:Distribution}
\end{figure}

To answer RQ3 and gain a better understanding of how developers deal with violation symptoms, we plot a tree map (see Figure~\ref{F:Distribution}) of the distribution of the status (i.e., ``\texttt{Merged}'', ``\texttt{Abandoned}'', and ``\texttt{Deferred}'') of the patches containing violation symptoms. We further analyzed the developers' reactions (including \textit{refactored}, \textit{removed}, and \textit{ignored}) in response to violation symptoms from code review comments. 

We found that most (76.1\%, 461 out of 606) of the violation symptoms are in ``\texttt{Merged}'' status, which means that developers agreed to merge the submitted code into the code repository. 23.1\% (140 out of 606) of the patches are in ``\texttt{Abandoned}'' status, which means that the submitted code was rejected to be integrated into the code repository. Only a few patches (0.8\%, 5 out of 606) stayed in ``\texttt{Deferred}'' status, which denotes a pending status and only exists in the code review of Qt (the reviewers and developers consider that the raised issues are not of very high priority and can be fixed in the following releases).

For the patches that contain violation symptoms, 77.7\% (358 out of 461) of the merged patches and 82.9\% (116 out of 140) of the abandoned patches, were addressed by refactoring. Moreover, 12.8\% (59 out of 461) and 9.5\% (44 out of 461) of violation symptoms were removed (i.e., deleted the code) and ignored (i.e., no changes), respectively, in the merged patches. Similar percentage of violation symptoms were removed (7.1\%) and ignored (10.0\%) in the abandoned patches. In the five deferred patches, developers refactored four submitted code snippets to cope with the violation symptoms and ignored one of them, while the remaining issues would be addressed in future releases.

\section{Discussion}\label{sec:Discussion}
In this section, we first interpret the study results, and we then discuss the implications of the results for practitioners and researchers.

\subsection{Interpretation of Results}\label{sec:Interpretation}

The percentage of the identified violation symptoms from code review comments is rather low (2.8\%, 606 out of 21,583) in the selected four OSS projects (i.e., Nova, Neutron, Qt Base, and Qt Creator). Prior studies \cite{li2022sae, Paixao2019icr} show that the percentage of architecturally-relevant information is comparatively lower than code-level issues (e.g., code smells) in code reviews, and our results comply with the findings from previous studies. Although the low percentage of architecturally-relevant code review comments, the identified architectural issues (especially architecture violation symptoms) can have a seriously negative impact on software maintenance and evolution. 

\textbf{RQ1: Categories of violation symptoms}. We classified the collected violation symptoms into three categories of violation symptoms with ten subcategories that developers often discuss during development. Design-related violations are the main category of violation symptoms. Specifically, we observed that structural inconsistency is the most common subcategory of violation symptoms. Structural inconsistency might be triggered by classes with many methods that make systems tend to be complex, overloaded, and contain architectural smells, and such classes have a greater possibility of being reused and becoming the source of architectural inconsistencies \cite{Lenhard2019ess}. Moreover, structural inconsistencies might be hard to detect with tools. For instance, \textit{architecture mismatch} is a kind of structural inconsistency that is relatively difficult to detect by off-the-shelf tools due to various reasons (e.g., standard architectural description languages are used to document architecture, but they generally do not support tool-assisted detection of architecture mismatches) \cite{Garlan2009archm}. 

Additionally, our results show that certain design-related factors are also common sources of violation symptoms that we cannot ignore. For instance, violations of layered pattern (i.e., \textit{violation of architecture patterns} in Section~\ref{sec:Results_RQ1}) undermine the sustainability and reliability of systems and may gradually lead to architecture erosion due to their accumulation. In many cases, such violations usually require considerable effort to repair, to the extent that such a repair may not be financially feasible \cite{li2022SMS, sarkar2009mlba}. Besides, our results indicate that certain design-related violations (e.g., \textit{violations of design decisions}, \textit{design principles}, and \textit{rules}) are prevalent during development, and one possible reason is that the missing architectural knowledge leading to these violations might exist in certain small groups such as architects or team leaders. Therefore, our results suggest that disseminating and sharing the architectural knowledge related to violation symptoms across the development team is necessary. 

\textbf{RQ2: Linguistic patterns expressing violation symptoms}. The results of RQ2 show that most (60\%) of violation symptoms contain the terms about ``\textit{inconsistent}'' (e.g., not consistent and inconsistency) and ``\textit{violate}'' (e.g., violation and violating) (see Figure~\ref{F:terms}). 
One possible explanation is that such terms are more in line with the idiomatic expressions used by developers and reviewers, and they are commonly used to discuss issues related to violation symptoms in system design.

Regarding the linguistic patterns, the results show that the linguistic patterns of expressing violation symptoms from code review comments can be mapped to the categories of linguistic patterns identified in development emails \cite{di2016deca}. The six linguistic patterns can also be used to analyze textual artifacts in other sources (e.g., app reviews \cite{di2016wwu, di2017ufb} and issue reports \cite{Huang2018aim}), and our findings indicate that code review discussions also encompass these categories of linguistic patterns. Besides, the major type of linguistic patterns of expressing violation symptoms is \textit{Problem Discovery} (see Table~\ref{T:Categories and Percentages}). We conjecture that developers incline to use the linguistic patterns regarding \textit{Problem Discovery} (see Table~\ref{T:linguistic patterns}) to specify violation symptoms, as one of the aims of code review is to identify issues during development.

\textbf{RQ3: Reactions to violation symptoms from developers}. We found that most of the identified violation symptoms were merged into the code repositories after refactoring or removing the smelly code. This observation indicates that code review is necessary, and to a large extent these review comments can help to mitigate the risk of architecture erosion caused by violation symptoms. In addition, developers' reactions (i.e., 77.7\% patches are merged into the code base) also indicate that the review comments raising violation symptoms are crucial, especially for large-scale and long-term projects \cite{Bosu2016pas}, as these review comments help reduce violation symptoms and improve the quality of software systems. 

In some sense, \textit{remove} the code that contains violation symptoms can also be considered as code improvement, for example, removing duplicated code or redundant dependencies decreases code complexity and increases system maintainability. Therefore, the percentage of \textit{improvement} (\textit{refactored} + \textit{removed}) regarding addressing violation symptoms accounts for around 90\% no matter whether it is merged into the code repositories or not. Only a small percentage (9.6\%) of the identified violation symptoms were ignored and remained in the systems, and one possible reason is that developers have different opinions about how to address the remaining violation symptoms without reaching an agreement. Another possible reason is that the submitted code containing violation symptoms is not quite urgent to be fixed and has a lower priority, as the priority of issues depends on the severity and degree of the impact on different quality attributes \cite{li2022SMS}. In general, the study results show that developers incline to repair the issues related to violation symptoms when they are pointed out or discussed during code review.

\subsection{Implications}\label{sec:Implications}

\subsubsection{Implications for researchers} 
We identified three categories of violation symptoms from code review comments. Researchers are encouraged to investigate violation symptoms from other artifacts (such as pull requests, issues, and developer mailing lists) to provide more comprehensive empirical evidence, in order to further validate and consolidate the observations in this study. For example, the categories of violation symptoms can be further explored because only four OSS projects (written in Python and C++) from two communities (OpenStack and Qt) were used in our study. It would be worth exploring violation symptoms in both industrial and OSS projects written in other programming languages (e.g., Java) and communities (e.g., Apache). Besides, researchers can further conduct a comparison between the identified violation symptoms by utilizing certain off-the-shelf architecture conformance checking techniques (e.g., reflexion models \cite{Murphy1995sfm}) and our dataset (i.e., manually collected violation symptoms). More specifically, they can try to perform quantitative comparisons regarding the identified violation symptoms between code and textual artifacts (e.g., code review comments and commit messages) with the purpose of evaluating the performance of the techniques. In addition, researchers can also try to map the violation symptoms from textual artifacts to code in order to improve and complement the existing architecture conformance checking techniques.

Moreover, we have created a dataset \cite{Replication} containing violation symptoms of architecture erosion from code review comments. This dataset can act as a foundation for future study on architecture erosion \cite{li2022SMS}, especially architecture violation symptoms. For example, researchers can further explore the possibility of automatic identification of violation symptoms from textual artifacts through employing natural language processing techniques based on machine learning and deep learning algorithms. Automatically identifying violation symptoms would be of great value to developers, as manual identification can be extremely tedious, effort-intensive, and error-prone. Specifically, based on the models trained by machine learning and deep learning algorithms, researchers can devise auxiliary plugins to existing code review tools for providing warnings of violation symptoms to developers during development and maintenance. 

\subsubsection{Implications for practitioners}
The results in Section~\ref{sec:Results} and their explanations in Section~\ref{sec:Interpretation} can be used by practitioners to guide their refactoring and maintenance activities. For example, having the categories of violation symptoms might help the practitioners to be aware of the possible violations in system architecture and then consider avoiding or repairing such issues in their daily development work. Moreover, the frequently-used terms and linguistic patterns related to descriptions of violation symptoms from the viewpoint of developers, can help developers pay more attention to violation symptoms during maintenance and evolution. For example, such terms and linguistic patterns could be a clear signal that developers should be wary of architecture erosion risks and avoid the appearance of violation symptoms. 

Furthermore, we encourage practitioners to manage violation symptoms with the purpose of facilitating refactoring and repairing architecture violations. As reported by Schultis \textit{et al}. in Siemens, architecture violations must be explicitly managed, which includes addressing the existing architecture violations and preventing future violations \cite{Schultis2016}. 
Therefore, architecture teams (especially architects) should take the responsibility for collecting and monitoring violation symptoms, and then equip developers with the knowledge to repair or minimize architecture violations during development. Thus, practitioners can work with researchers and put effort to developing dedicated tools for managing violation symptoms and improving the productivity of maintenance activity.

\section{Threats to Validity}\label{sec:Threats}
The threats to the validity of this study are discussed by following the guidelines proposed by Wohlin \textit{et al}. \cite{Wohlin2012ese}. Internal validity is not considered because this study does not address any causal relationships between variables.

\textbf{Construct validity} pertains to whether the theoretical and conceptual constructs are correctly interpreted and measured. In this work, one potential threat is about the construction of the keyword set. To mitigate this threat, we first built the keyword set based on previous studies, and then we employed a pre-trained word embedding model to query and select similar keywords in the software domain. Besides, we constructed and used the co-occurrence matrix approach proposed by Bosu \textit{et al}. \cite{Bosu2014icc} to check the possible missing co-occurring words. In this way, the potential threat can be, at least partly, mitigated.

\textbf{External validity} concerns the extent to which we can generalize the findings to other studies. First, a potential threat to external validity is whether the selected projects are representative enough. Our work chose the two largest and most popular OSS projects (i.e., Nova and Neutron) from the OpenStack community, both written in Python language; we further selected another two major OSS projects (i.e., Qt Base and Qt creator) from the Qt community, which are written in C++. Second, another threat is that only Python and C++ OSS projects were selected, which may reduce the generalizability of the study results. Our findings may not generalize or represent all open source and closed source projects. It would be interesting to select more projects from different sources and programming languages to increase the external validity of the study results. Besides, it is worth exploring the generalizability of the findings regarding the frequent terms and linguistic patterns related to the identified violation symptoms in this work, for example, to investigate whether these findings are also applicable with other artifacts (e.g., pull requests and issues).

\textbf{Reliability} refers to the replicability of a study regarding yielding the same or similar results when other researchers reproduce this study. The potential threat is mainly from the activities of data collection and data analysis. For data collection, we presented the detailed data collection steps in Section~\ref{sec:Data Collection} and provided a replication package \cite{Replication} for reproducing the data collection and filtering process, which can help to enhance the reliability of the results. Regarding data labeling, our observations show that developers generally discussed individual violation symptoms within one single review comment, as such, the threat of multiple symptoms discussed in one review comment with multiple labels is not present in this work. As for the data analysis, to mitigate personal bias, we conducted a pilot labeling and classification (see Phase I in Section~\ref{sec:Data Analysis}) before the formal data labeling and classification process, and we got a Cohen's Kappa value of 0.857, which indicates a substantial inter-rater agreement. Likewise, we executed a similar process (see Phase II in Section~\ref{sec:Data Analysis}) when we conducted the formal data labeling and analysis. Any disagreements were discussed between the four researchers to reach an agreement and at least two researchers participated in the data labeling and classification process.

\section{Related Work}\label{sec:Related Work}
In this section, we discuss the work related to our study, which involves architecture violations and their detection approaches (i.e., architecture conformance checking), as well as the data sources (i.e., code review comments) used in this study.

\subsection{Architecture Violations}\label{sec:Violations}
Over the past decades, there have been extensive investigation on architecture violations. Brunet \textit{et al}. \cite{Brunet2012enav} performed a longitudinal study to explore the evolution of architecture violations in 19 bi-weekly versions of four open source systems. They investigated the life cycle and location of architecture violations over time by comparing the intended and recovered architectures of a system. They found that architecture violations tend to intensify as software evolves and a few design entities are responsible for the majority of violations. More interestingly, some violations seem to be recurring after being eliminated. Mendoza \textit{et al}. \cite{Mendoza2021avd} proposed a tool ArchVID based on model-driven engineering techniques for identifying architecture violations, and the tool supports recovering and visualizing the implemented architecture. 

Moreover, Terra \textit{et al}. \cite{Terra2015rsr} reported their experience in fixing architecture violations. They proposed a recommendation system that provides refactoring guidelines for developers and maintainers to repair architecture violations in the module architecture view of object-oriented systems. The results show that their approach can trigger correct recommendations for 79\% architecture violations, which were accepted by architects. Maffort \textit{et al}. \cite{Maffort2016mav} proposed an approach to check architecture conformance for detecting architecture violations based on defined heuristics. They claimed that their approach relies on the defined heuristic rules and can rapidly raise architectural violation warnings. Different from the abovementioned studies focusing on detecting architecture violations in source code, our work investigates the architectural violation symptoms in code review comments from the perspective of developers, including the categories and linguistic patterns of expressing violation symptoms, as well as the reactions developers take to deal with violation symptoms.


\subsection{Architecture Conformance Checking}\label{sec:Architecture Conformance Checking}
Architecture conformance checking techniques are the most commonly-used approaches to detect architecture violations \cite{li2022SMS}. They can be checked statically or dynamically, and they are usually performed to compare the structure of the intended architecture (provided by the architects) with the extracted architecture information from source code that implements the architecture. For example, Pruijt \textit{et al}. \cite{Pruijt2014srm} proposed a metamodel for extensive support of semantically rich modular architectures in the context of architecture conformance checking. Miranda \textit{et al}. \cite{Miranda2016acc} presented an architectural conformance and visualization approach based on static code analysis techniques and a lightweight type propagation heuristic. They evaluated their approach in three real-world systems and 28 OSS systems to identify architecture violations.

Besides, rule-based conformance checking approaches are also employed to identify architecture violations. For example, previous studies detected architecture violations by checking the explicitly defined architectural rules \cite{Mendoza2021avd, Schroder2017acc, Terra2009dcl}. Moreover, it is viable to check architecture conformance and identify architecture violations by defining and describing the systems through Architecture Description Languages (ADLs) \cite{Terra2009dcl, Rocha2017DCL}, or Domain-Specific Languages (DSLs) \cite{Caracciolo2015uaa, Juarez2017pee}. However, the aforementioned approaches have obvious limitations; for example, much effort is required to address the challenges of understanding the architecture design (e.g., concepts and relations), defining architectural rules (or description languages) in advance, and establishing a mapping between architectural elements and source code. Moreover, other limitations, such as lack of generalizability, visualization of architecture views, and insufficient tooling support, hinder the above approaches from being widely used in practice. Additionally, to the best of our knowledge, prior studies regarding architecture violations focus on checking architecture conformance with source code using predefined abstract models and rules, and there is no evidence-based knowledge on identifying architecture violations from textual artifacts, such as code review comments.

\subsection{Code Review Comments}\label{sec:Code Review Comments}
Code review comments contain massive knowledge related to software development, and a variety of studies analyzed software defects and evolution through mining review comments and commit records. 
Zhou and Sharma \cite{Zhou2017ais} designed an automated vulnerability identification system based on a large number of commits and bug reports (containing rich contextual information for security research), and their approach can identify a wide range of vulnerabilities and significantly reduce false positives by more than 90\% compared to manual effort. 

Besides, Uchôa \textit{et al}. \cite{Uchoa2020hdm} investigated the impact of code review on the evolution of design degradation through mining and analyzing a plethora of code reviews from seven OSS projects. They found that there is a wide fluctuation of design degradation during the revisions of certain code reviews. Paixão \textit{et al}. \cite{Paixao2020bti} explored how developers perform refactorings in code review, and they found that refactoring operations are most often used in code reviews that implement new features. Besides, they observed that the refactoring operations were rarely refined or undone along the code review, and such refactorings often contribute to new code smells and bugs. Given that previous studies discussed above investigated various aspects of code review regarding development and maintenance (e.g., decisions and design degradation), there are no studies that investigate architecture violations through code review comments; we decided to explore the violation-related issues (i.e., violation symptoms) from code review comments.

\section{Conclusions}\label{sec:Conclusions}
As software systems evolve, the changes in the systems could lead to cascading violations, and consequently the architecture will exhibit an eroding tendency. In this work, we conducted an empirical study to investigate the discussions on violation symptoms of architecture erosion from code review comments. We collected a large number of code review comments from four popular OSS projects in the OpenStack (i.e., Nova and Neutron) and Qt (i.e., Qt Base and Qt Creator) communities. Our results show that ten subcategories of violation symptoms in three main categories are discussed by developers during the code review process. Besides, we found that the most frequently-used terms related to the description of violation symptoms concern \textit{structural inconsistencies}, \textit{design-related violations}, and \textit{implementation-related violations}, such as \textit{violation of design decisions}, \textit{design principles}, and \textit{architecture patterns}; the most common linguistic pattern (90.4\%) used to express violation symptoms is \textit{Problem Discovery}. \textit{Refactoring} is the major measure that developers used to address violation symptoms, no matter whether the smelly code is integrated (i.e., 77.7\% refactorings happened in the merged patches) or not (i.e., 82.9\% refactorings happened in the abandoned patches). The finding indicates that code review can help reduce violation symptoms and increase system quality.

Our findings encourage researchers to investigate violation symptoms from various artifacts (e.g., pull requests, issues, and developer mailing lists) in order to provide more comprehensive evidence for validating and consolidating the findings. The most frequently-used terms and linguistic patterns used to express violation symptoms can help researchers and practitioners better understand and be aware of the natural language on describing violation symptoms of architecture erosion commonly used by developers. Besides, explicitly managing violation symptoms can to some extent help reduce the occurrence of architecture violations and prevent future violations during development and maintenance.

Developers usually discuss and address design-related issues in artifacts such as commits, issues, and pull requests \cite{Brunet2014ddd}. In this context, we plan to construct classification models based on textual artifacts with machine learning and deep learning techniques for the purpose of automatically notifying developers about the potential violation symptoms of architecture erosion during development; for example, as a plugin to the Gerrit tool during the code review process. We also plan to invite practitioners to evaluate the effectiveness and efficiency of the proposed classification models and the tool on assisting developers in detecting violation symptoms.

\section*{\textbf{Acknowledgements}}
This work has been partially supported by the National Natural Science Foundation of China (NSFC) with Grant No. 62172311 and the Special Fund of Hubei Luojia Laboratory.

\balance
\bibliographystyle{elsarticle-num}
\bibliography{ref}
\balance
\end{document}